\pdfoutput=1
\documentclass[runningheads]{llncs}

\usepackage{graphicx}
\graphicspath{{./Images/}}
\usepackage{float}
\usepackage{hyperref}               % for URL formatting

\usepackage{caption}
\usepackage{subcaption}
\usepackage[export]{adjustbox}

\usepackage{multirow}
\usepackage{tabularx}
\usepackage{longtable}
\usepackage[table,xcdraw]{xcolor}
\usepackage[normalem]{ulem}
\useunder{\uline}{\ul}{}

\usepackage{cite}
\begin{document}
\title{Web Archiving as Entertainment}
\author{Travis Reid\inst{1}\orcidID{0000-0003-1360-7963} \and
Michael L. Nelson\inst{1}\orcidID{0000-0003-3749-8116} \and
Michele C. Weigle\inst{1}\orcidID{0000-0002-2787-7166}}
\authorrunning{T. Reid et al.}
\institute{Old Dominion University, Norfolk VA 23529, USA\\
\email{treid003@odu.edu, \{mln, mweigle\}@cs.odu.edu}}
\maketitle              % typeset the header of the contribution
\begin{abstract}
We want to make web archiving entertaining so that it can be enjoyed like a spectator sport. To this end, we have been working on a proof of concept that involves gamification of the web archiving process and integrating video games and web archiving. Our vision for this proof of concept involves a web archiving live stream and a gaming live stream. We are creating web archiving live streams that make the web archiving process more transparent to viewers by live streaming the web archiving and replay sessions to video game live streaming platforms like Twitch, Facebook Gaming, and YouTube. We also want to live stream gameplay from games where the gameplay is influenced by web archiving and replay performance. So far we have created web archiving live streams that show the web archiving and replay sessions for two web archive crawlers and gaming live streams that show gameplay influenced by the web archiving performance from the web archiving live stream. We have also applied the gaming concept of speedruns, where a player attempts to complete a game as quickly as possible. This could make a web archiving live stream more entertaining, because we can have a competition between two crawlers to see which crawler is faster at archiving a set of URIs.

\end{abstract}
%
%
%
%\begin{figure}[htbp] \centering \includegraphics[scale=0.1333]{Images/1080p_fit_Gaming_vs_Archiving_vs_Sports_Participating.jpg} \caption{Different ways to participate in gaming (left), web archiving (center), and sport sessions (right)} \label{fig:gamesVsArchivingvsSports} \end{figure}

\section{Introduction} \label{Introduction}
Game walkthroughs are guides that show viewers the steps the player would take while playing a video game. Recording and streaming a user's interactive web browsing session is similar to a game walkthrough, because it shows the steps the user would take while browsing different websites. The idea of having game walkthroughs for web archiving was first explored in 2013 \cite{mln-blog13}, but the web archive crawlers at that time were not ideal for web archiving walkthroughs because they did not allow the user to view the web page as it was being archived. Recent advancements in web archive crawlers have made it possible to preserve the experience of dynamic web pages by recording a user's interactive web browsing session, which makes it possible to create a walkthrough of a web archiving session.

\begin{figure}[htbp] \centering \includegraphics[width=\linewidth]{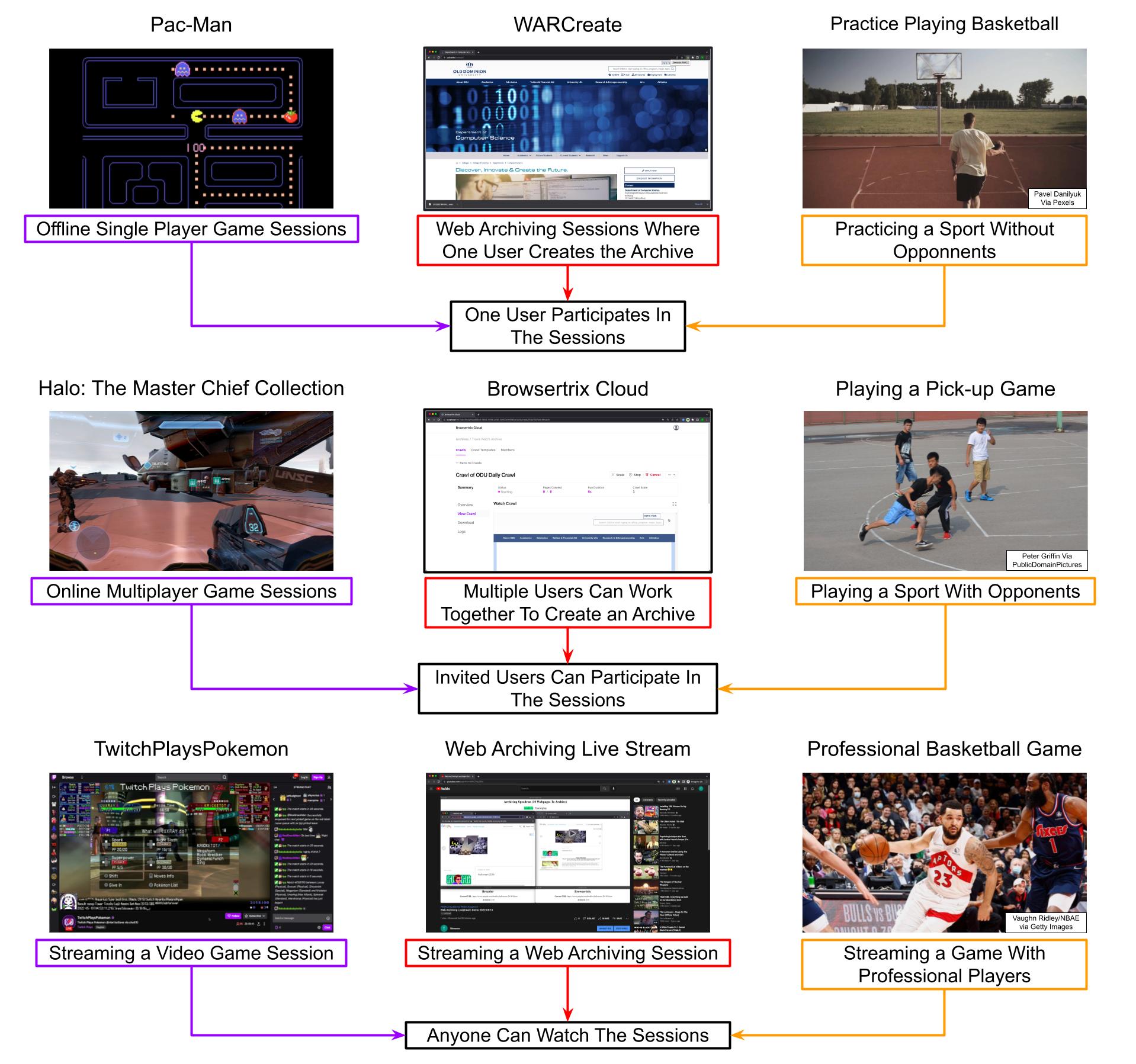} \caption{Different ways to participate in gaming (left), web archiving (center), and sport sessions (right)} \label{fig:gamesVsArchivingvsSports} \end{figure}

Figure \ref{fig:gamesVsArchivingvsSports} applies the analogy of different types of video games and basketball scenarios to types of web archiving sessions. Practicing playing a sport like basketball by yourself, playing an offline single-player game like Pac-Man, and archiving a web page with a browser extension like WARCreate \cite{warcreate-jcdl2012} have similar qualities, because only one user or player is participating in the session (Figure \ref{fig:gamesVsArchivingvsSports}, top row). Playing team sports with a group of people, playing an online multiplayer game like Halo, and collaboratively archiving web pages with Browsertrix Cloud \cite{kreymer-BrowsertrixCloud-gh21} are similar since multiple invited users or players can participate in the sessions (Figure \ref{fig:gamesVsArchivingvsSports}, center row). Watching a professional sport on ESPN+, streaming a video game on Twitch, and streaming a web archiving session on YouTube can be similar because anyone can be a spectator and watch the sporting event, gameplay, or web archiving session (Figure \ref{fig:gamesVsArchivingvsSports}, bottom row).

One of our goals in the Game Walkthroughs and Web Archiving project \cite{gameWalkthoughs-iipc22} is to create a web archiving live stream like that shown in Figure \ref{fig:gamesVsArchivingvsSports}. We want to create a web archiving live stream that is entertaining so that it can be enjoyed like a spectator sport. To this end, we have been working on a proof of concept that involves integrating video games with web archiving and the gamification of the web archiving process. A part of the vision for this proof of concept is to create web archiving live streams that make the web archiving process more transparent to viewers by live streaming the web archiving and replay sessions. Another part of the vision is to create gaming live streams that show viewers gameplay that is influenced by web archiving and replay performance. We have completed some of the goals for this vision. We have created automated web archiving live streams \cite{archivingSpeedrunWithReplay-yt22} where the web archiving process has been made transparent by live streaming the crawling of web pages and the playback of the archived web pages. The web archiving process is also gamified during our web archiving live streams by applying the gaming concept of a speedrun to the web archiving process. For video games, a speedrun is a playthrough of a game where the player attempts to complete a section of the game or the entire game as quickly as possible \cite{wiki-speedrun}. Applying this concept to web archiving could make the web archiving live stream more entertaining, because we can have a competition between two crawlers to see which crawler is faster at archiving a set of URIs. We have also created automated gaming live streams \cite{gamingPlaylist-yt22} where the capabilities for the in-game characters were determined by the web archiving performance from a web archiving live stream.

\section{Background}
This section covers the background information for the tools that we used and the trends in gaming and web archiving that made it possible to work on this proof of concept.

\subsection{Web Archive Crawlers and Replay Systems} \label{Web Archive Crawler and Replay Systems}
A web crawler is a program that can download web pages, extract hyperlinks from web pages, and recursively download the web pages associated with the extracted hyperlinks \cite{najork09}. Before downloading web pages, a web crawler is given a set of seed URIs. A web archive crawler saves the captured web content into an archive. The archived version of a web page is called a memento. Traditional archival crawlers like Heritrix\cite{heritrix-gh11} do not use a web browser when archiving a web page and cannot perform interactions with a web browser by itself. There are also browser-based archiving tools that can use a headless browser, a regular browser like Chrome, or the current user's browser. A headless browser is a web browser that can render web pages and execute JavaScript, but does not use a graphical user interface (GUI). Some browser-based archiving tools like Squidwarc\cite{berlin-gh17}, Brozzler\cite{Levitt-brozzler-gh14}, and Browsertrix Crawler\cite{kreymer-BrowsertrixCrawler-gh20} have options to use a headless web browser or a regular web browser. Web archiving tools like WARCreate\cite{warcreate-jcdl2012} and ArchiveWeb.page\cite{kreymer-ArchiveWebpage-gh20} use the current user's browser, because they are browser extensions. Some example cases that affect the archiving performance of a web archive crawler are web pages that dynamically load content, require the use of server-side events, or require user account authentication.

A web archive replay system is used to view archived content. Wayback\cite{Wayback-gh12}, OpenWayback\cite{OpenWayback-gh12}, pywb\cite{pywb-gh13}, InterPlanetary Wayback (ipwb)\cite{alam-jcdl16}, and ReplayWeb.page\cite{ReplayWebPage-gh20} are examples of web archive replay systems. Some examples of what can affect the performance of a replay system are the number of users making a replay request at the same time and the number of requests made when loading the archived resources. If a replay system does not correctly rewrite the URIs for resources, then it may result in live resources being loaded instead of the expected archived resource. Lerner et al. \cite{lerner-ccs17} and Brunelle \cite{brunelle-zombies-wsdl12} have shown examples where the URIs used by the JavaScript for a web page have caused issues with URL rewriting and lead to leakage in the archived website.

\subsection{Trends in Gaming and Web Archiving} \label{Trends in Gaming and Web Archiving}
Gaming is a popular hobby, and over the past twenty years esports (gaming competitions) have become a popular spectator sport that has tens of millions of viewers each year \cite{hollist-alr15}. This growth has also been reflected in the number of esports tournaments. In 2000, there were only 65 tournaments, but by 2020 there were almost 5000 tournaments held\cite{esportsearnings22}. Since gaming is a popular form of entertainment we wanted to integrate gaming with web archiving so that web archiving can become more entertaining to watch.

Recent advancements that have been made for web archive crawlers make it possible for us to work on this proof of concept (integrating gaming, web archiving, and live streaming). Traditional web archive crawlers do not use a web browser when archiving a web page. This prevents viewers of a web archiving live stream from being able to see what the crawler is currently archiving and the interactions that the crawler is performing on the web page. Now, we have browser-based web archiving tools like Squidwarc, Brozzler, Browsertrix Crawler, and ArchiveWeb.page that allow the user to view a web page while it is being archived. These tools have features that allow the user to view some of the automated interactions that are done while the crawlers are archiving the web page like scrolling down the web page or moving to different fragments on a web page. Browser-based crawlers are important for the web archiving live streams, because the content being shown during the live stream will be the web archiving and replay sessions.

\subsection{App Automation Tool} \label{App Automation Tool}
GUI testing tools are used to automate tests for applications by simulating interactions like clicks and key presses. Selenium\cite{selenium} is an example of a browser automation tool that can perform automated interactions with a web browser so that a developer can test their web apps. Appium\cite{appium} is a tool that can be used to automate tests for desktop applications, mobile applications, and web applications. We have used both of these tools to customize gameplay based on the web archiving crawler performance.

\section{Related Work}
For this section we discuss related works that have either measured the performance of web crawlers, allowed multiple users to view the web archiving process for a browser-based crawler, automated gameplay, or used gamification to improve a non-gaming activity.

\subsection{Measuring Web Crawler Performance} \label{Measuring Web Crawler Performance}
One way to measure the performance of a crawler is to use the speed of the crawler, which can be measured by the number of web pages downloaded per second \cite{shkapenyuk-icde02, najork2002high}. When evaluating Google’s crawler and the Internet Archive’s crawler, Heydon and Najork \cite{heydon-www99} used the crawlers’ speed, the number of kilobytes per second, and the number of HTTP requests per day. Mohr et al. \cite{mohr-iwaw04} used the time it took for a crawl to finish and the number of URIs discovered per second to determine the performance of their crawler.

The performance of a web archive crawler can also be measured by how well the crawler archived a web page based on the quality of the replayed web page. Gray and Martin \cite{gray-dlib13} determined the quality of a memento based on the number of missing embedded resources. Brunelle et al. \cite{brunelle-jcdl15, brunelle-odu16} created an algorithm for measuring memento damage, which determines the quality of an archived web page by calculating the importance of each embedded resource instead of just counting the missing resources. They assign a weight to each resource, and the sum of the weighted importance for the missing embedded resources is used to determine the damage that had been done to the memento. Kiesel et al. \cite{kiesel-acm18} used machine learning to predict the quality of a memento from the differences between the screenshots of the original web page and the archived web page. Banos and Manolopoulos \cite{Banos_clear, Banos-ijdl15} created the CLEAR and CLEAR+ methods for determining the archivability of a website. The CLEAR+ method determined the potential for a website to be archived completely and accurately based on four facets: accessibility, standards compliance, cohesion, and use of metadata. Reyes Ayala \cite{Ayala-ijdl2021} used a grounded theory approach to assess how humans evaluate the quality of a web archive capture.

\subsection{Viewing the Web Archiving Process} \label{Viewing The Web Archiving Process}
Most of the browser-based web archiving tools mentioned in Section \ref{Trends in Gaming and Web Archiving} can be used to view the web archiving process, but the browser-based web archive system that is most related to our work is Browsertrix Cloud \cite{kreymer-BrowsertrixCloud-gh21}. Browsertrix Cloud is a crawling system that uses Browsertrix Crawler for each crawl. Browsertrix Cloud allows collaborative archiving where multiple users can work together to create a collection \cite{browsertrixCloud-faq}. This crawling system is similar to our work, because it allows multiple users to view the web archiving sessions for the web pages that are currently being archived. A web archiving live stream would also show the web archiving sessions for web pages that will be captured and added to a collection.

\subsection{Automating Gameplay} \label{Automating Gameplay}
Two related cases for automating a game are for testing a game's features and automating the interactions in a game based on the choices made by a chat during a live stream. Some game developers like Rare \cite{rare-gdc19} have used automated game testing to automate certain gameplay scenarios so that they can detect issues in a video game before the human players test the game. This is related to our work, because we will use tools to automate the gameplay for our gaming live streams. TwitchPlaysPokemon is a Twitch streamer that is a bot that reads the chat of its live stream and then performs those interactions in the game. This is related to our work, because it involves using automated interactions with an app (the Pokemon emulator) in order to play the game. Other Twitch Plays accounts were created because of the success of TwitchPlaysPokemon, and these other accounts played games from different genres, like a first person shooter game named Halo Combat Evolved\cite{emilygera-halo-polygon15}, a sports game QWOP\cite{hernandez-qwop-kotaku14} that is played in the browser using only the Q, W, O, and P keys, fighting games like Street Fighter\cite{hernandez-qwop-kotaku14}, and a difficult action role-playing game like Dark Souls \cite{walton-darkSouls-arstechnica14}.

\subsection{Using Gamification To Improve Non-Gaming Activities}
Gamification is “the use of game design elements in non-gaming contexts” \cite{deterding-mindtrek11}. Gamification has been used in different areas like education \cite{barata-icgdra13}, commerce \cite{hamari-ecra13}, data gathering \cite{guin-ijmr12}, e-health \cite{allam-jmir15}, and exercise \cite{hamari-ecis13}. A literature review by Hamari et al. \cite{hamari-hicss14} found that studies that used gamification for education/learning contexts had positive outcomes like increased motivation and engagement in the learning tasks. Their literature review covered a range of different contexts where gamification was applied like intra-organizational systems, sustainable consumption, and work. Most of the reviewed studies that were in this literature review reported that using gamification did produce positive effects and benefits. Using gamification to improve the experience of a non-gaming activity is related to our work, because we are going to gamify our web archiving live streams by applying gaming concepts to the web archiving process.

\section{Integrating Web Archiving, Gaming, and Live Streaming} \label{Integrating Web Archiving, Gaming, and Live Streaming}
%\begin{figure}[htbp] \centering \includegraphics[scale=0.125]{Images/Overview_Figure_With_Replay_Mode_Vertical_Condensed.jpg} \caption{The current process for running our web archiving live stream and gaming live stream} \label{fig:overview} \end{figure}
\begin{figure}[htbp] \centering \includegraphics[width=\linewidth]{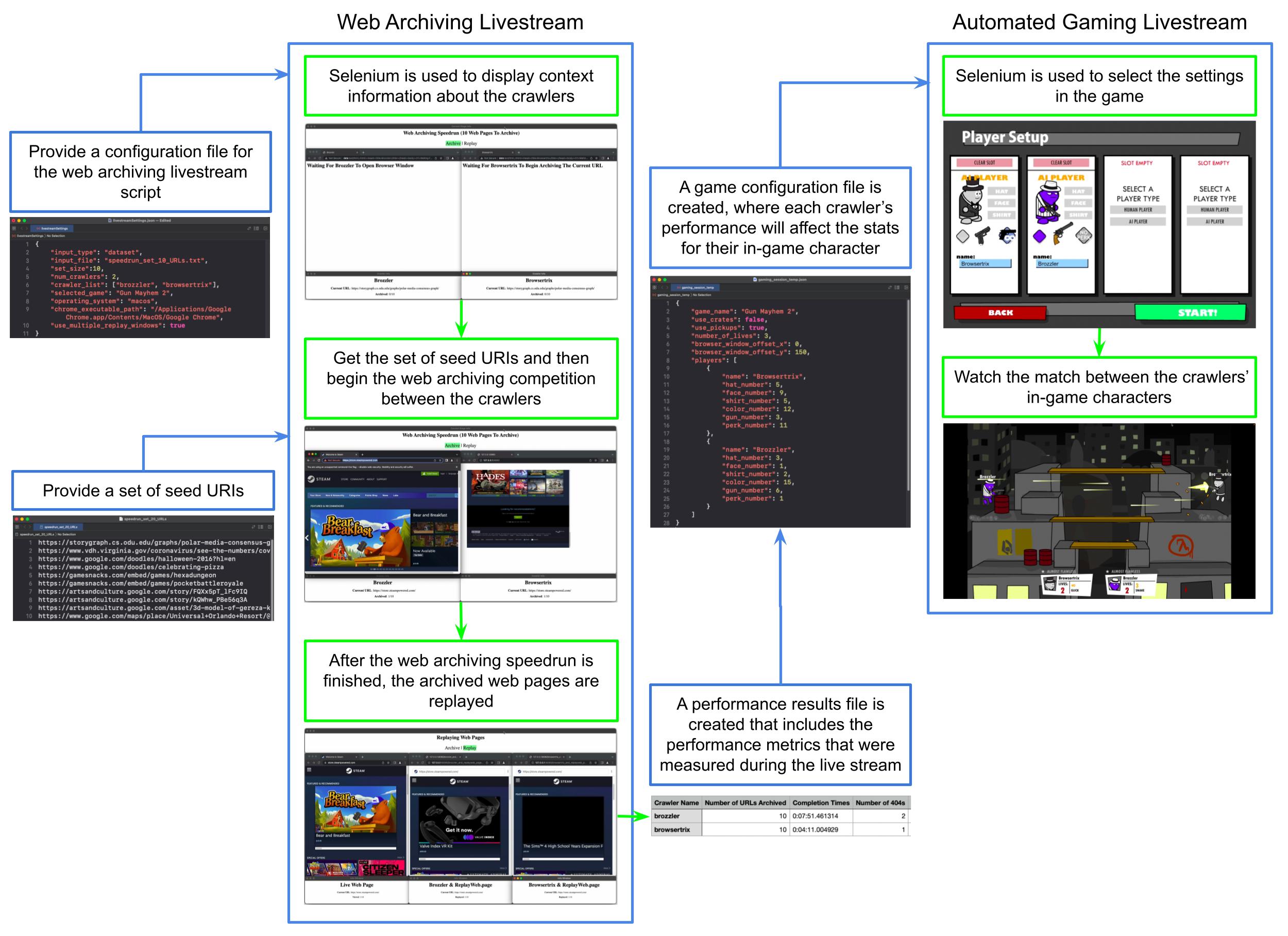} \caption{The current process for running our web archiving live stream and gaming live stream} \label{fig:overview} \end{figure}
%\begin{figure}[htbp] \centering \includegraphics[width=\linewidth]{Images/Overview_Figure_2022_4k_v3.jpg} \caption{The current process for running our web archiving live stream and gaming live stream} \label{fig:overview} \end{figure}
%
We currently have two kinds of live streams: a web archiving live stream and a gaming live stream. For the web archiving live streams, we want to make the web archiving process more transparent to viewers by streaming the archiving and replay sessions. For the gaming live streams, we want to show viewers gameplay from their favorite game where the gameplay is influenced by archiving and replay performance. Figure \ref{fig:overview} shows the current process that we are using for the live streams.

\begin{figure}[htbp]
    \centering
    \includegraphics[width=\textwidth]{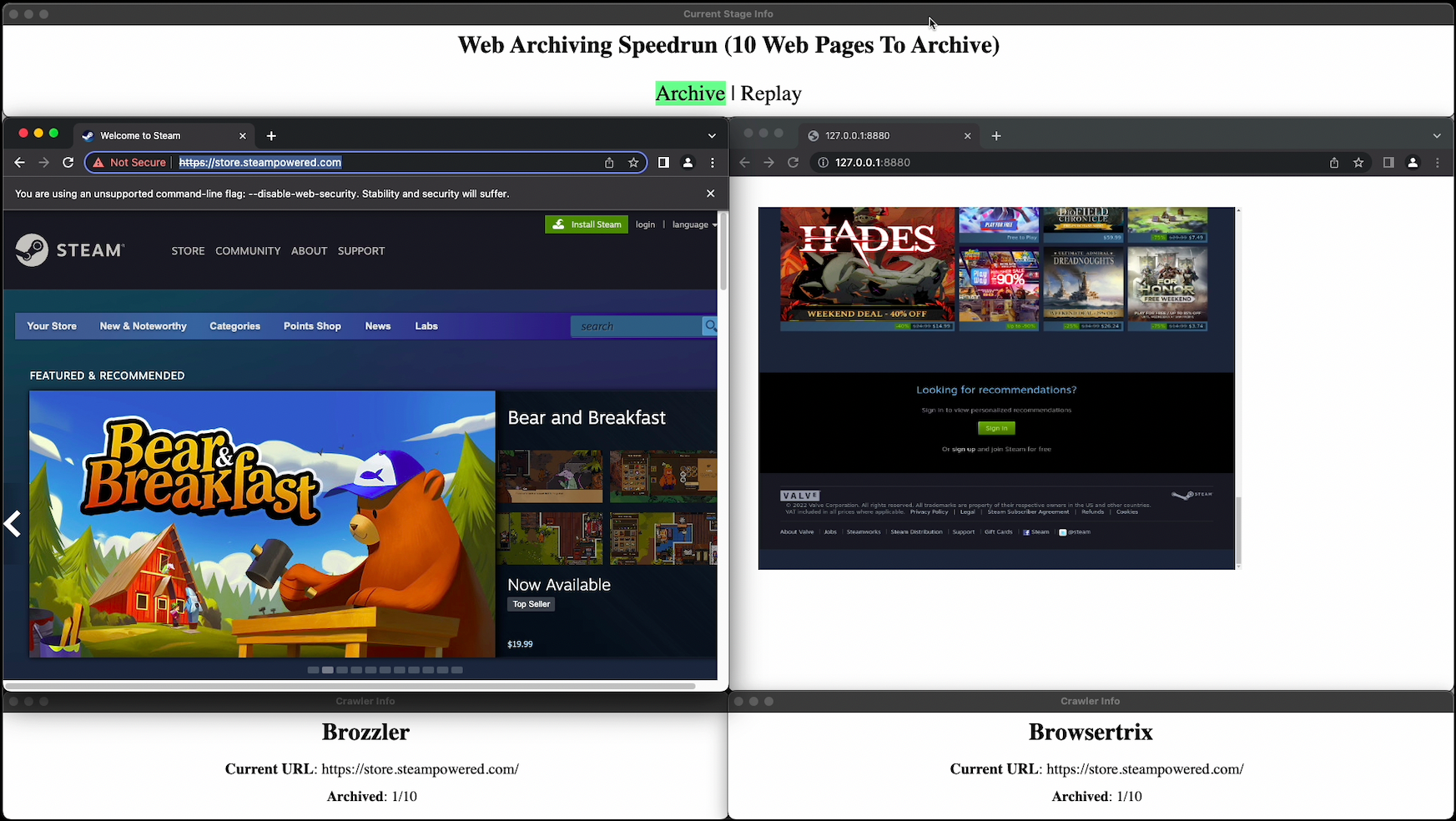}
    \caption{Viewers can watch crawlers archive web pages during our web archiving live streams}
    \label{fig:archiveMode}
\end{figure}

\begin{figure}[htbp]
    \centering
    \includegraphics[width=\textwidth]{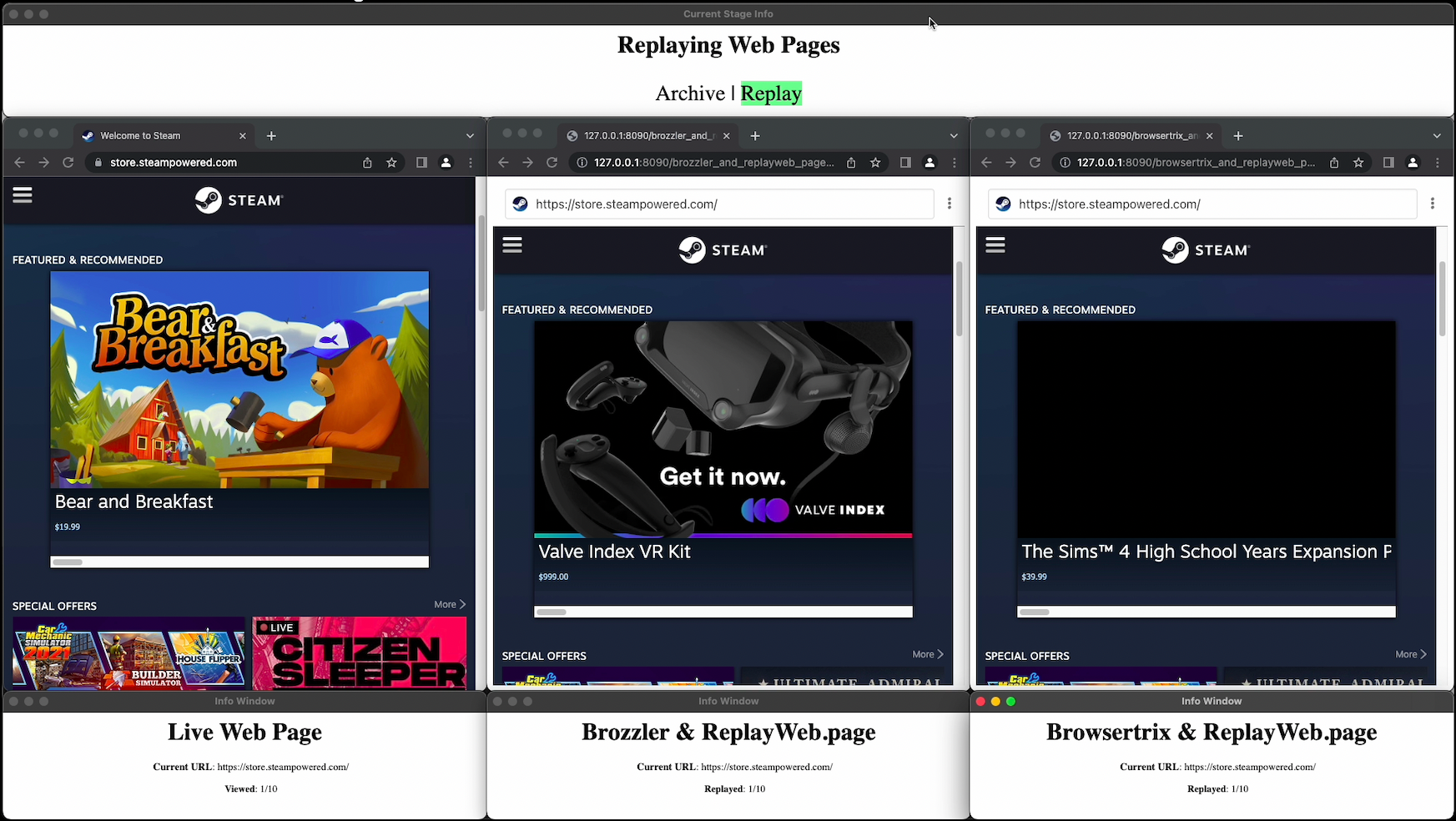}
    \caption{Replay sessions can be shown to viewers during our web archiving live streams}
    \label{fig:replayMode}
\end{figure}

For the web archiving live streams (Figure \ref{fig:overview}, left side), the viewers watch browser-based web crawlers archive web pages and the replay of the archived web pages. To make the live streams more entertaining, we made each web archiving live stream into a competition between crawlers to see which crawler performs better at archiving the set of seed URIs. The first step for the web archiving live stream is to use Selenium to setup the browsers that will be used during the live stream. The automated browsers are used to show information needed for the live stream, like the name and current progress for each crawler. The information currently displayed for a crawler’s progress is the current URL being archived and the number of web pages archived so far. The next step is to get a set of seed URIs that will be used for the competition and then let each crawler start archiving the URIs. During this step, the crawling of the web pages is shown to the viewers, which allows the viewers to observe the web archiving process in action (Figure \ref{fig:archiveMode}). Since browser-based crawlers are used during the live stream, the viewers can observe the interactions that the crawlers perform while archiving a web page like scrolling down the web page and switching between different URI fragments. The next step in the web archiving live stream is to replay the archived web pages after the crawlers are finished archiving the set of URIs. When replaying the archived web pages the live web page is shown beside the replayed web page so the viewers can see how well the crawler archived the web page (Figure \ref{fig:replayMode}). Showing the replay sessions allows the viewer to see which web pages are difficult to archive and if any resources are missing on the archived web page. ReplayWeb.page is the replay system that we are currently using for replay sessions. In future live streams we will also use other replay systems like pywb. At the end of the web archiving live stream a performance results file is created that includes the web archiving and replay performance metrics that were measured during the live stream. Currently this performance result file \cite{resultsFile-gh22} includes the number of web pages archived by the crawler during the competition, the speedrun completion time for the crawler, the number of resources in the WARC file with an HTTP response status code of 404, the number of resources with a different 400 or 500 level HTTP response status code, and the number of missing resources categorized by the file type (JavaScript, CSS, image, video, audio, and text/html). During future web archiving live streams, it will be possible for viewers to observe the results from the crawling session, because we will show a summary of what is included in the performance results file. Usually third parties cannot observe the results from a crawl when viewing a collection created by Archive-It users or a collection from the Internet Archive’s web collections unless the web collection has a CDX summary (example collection with a CDX summary \cite{barreau-ia22}) or a CDX file associated with the collection that can be downloaded and used with a CDX summary tool like CDX Summary \cite{CDXSummary-gh22} or cdx-summarize \cite{maurer-iipc22, cdxsummarize-gh22}.

\begin{figure}[htbp]
    \centering
    \includegraphics[width=\textwidth]{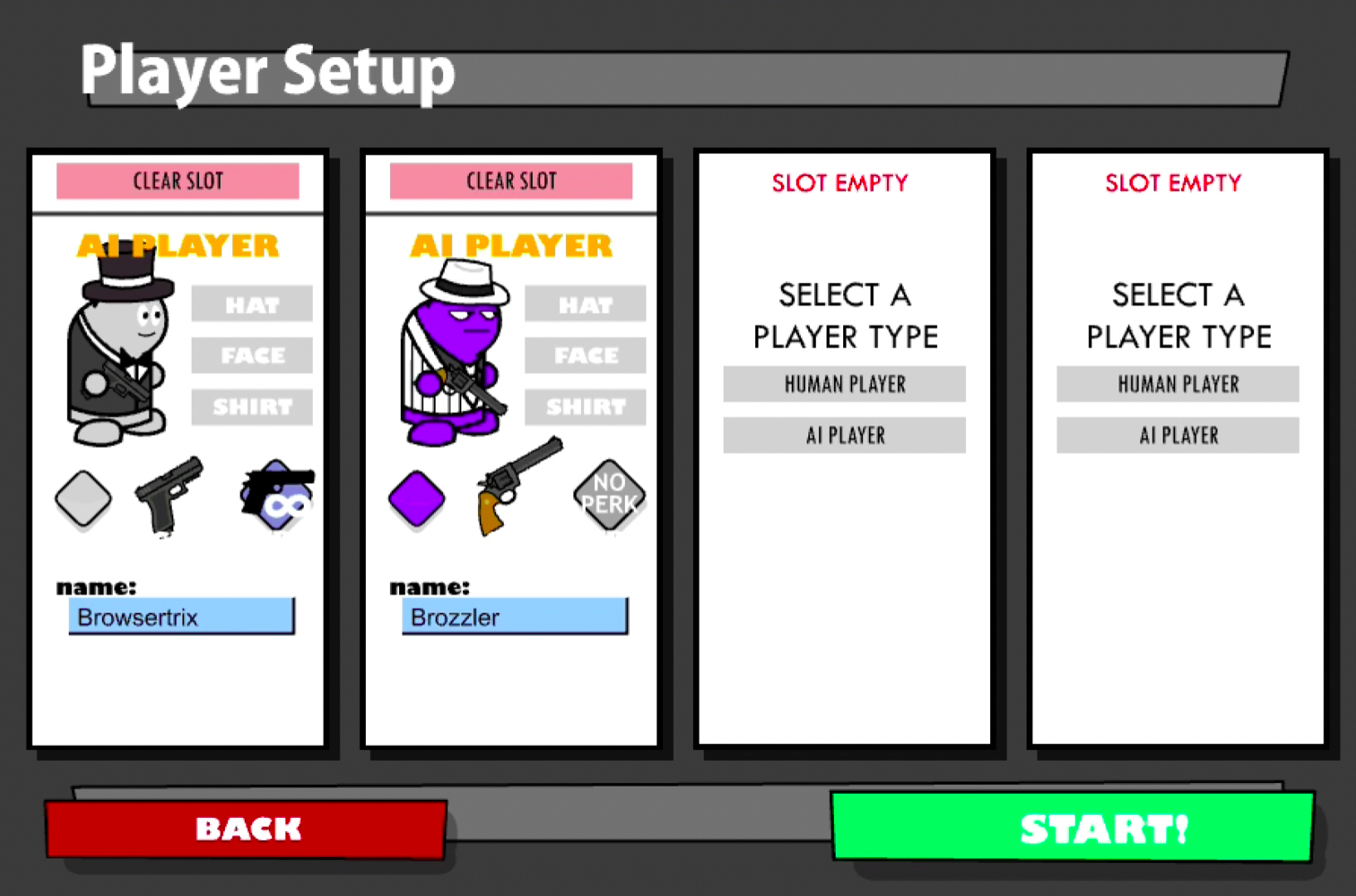}
    \caption{Using Selenium to customize in-game characters for Gun Mayhem 2 More Mayhem based on web archiving performance}
    \label{fig:gameSettings}
\end{figure}

The automated gaming live stream (Figure \ref{fig:overview}, right side) was created so that viewers can watch a game that has gameplay influenced by the web archiving and replay performance results from a web archiving live stream or any crawling session. Watching the gameplay is another way to view how well the crawlers and replay system performed during the archiving and replay sessions. This should be more entertaining to watch than viewing a summary of the performance results, especially when applied to long-running crawls, such as the British Library's domain crawl which can take up to three months \cite{BL:domaincrawl:2020}. Before an in-game match starts, a game configuration file is needed since it contains information about the selections that will be made in the game for the settings. The game configuration file is modified based on how well the crawlers performed during the web archiving live stream. If a crawler had good performance during the web archiving live stream, then the in-game character associated with the crawler will have better items, perks, and other traits. If a crawler performs poorly, then their in-game character will have the worst character traits. At the beginning of the gaming live stream, an app automation tool like Selenium (for browser games) or Appium (for locally installed PC games) is used to select the settings for the in-game characters based on the performance of the web crawlers (Figure \ref{fig:gameSettings}). After the settings are selected by the app automation tool, the match is started and the viewers of the live stream can watch the match between the crawlers’ in-game characters. We have initially implemented this process for two video games, Gun Mayhem 2 More Mayhem and NFL Challenge, however any game with a mode that does not require a human player and that allows character traits or items for each player to be assigned to different values could be used for an automated gaming live stream.

\section{Gamification of the Web Archiving Process} \label{Applying a Game Concept to the Web Archiving Process}
\begin{figure}[htbp]
    \centering
    \includegraphics[scale=0.175]{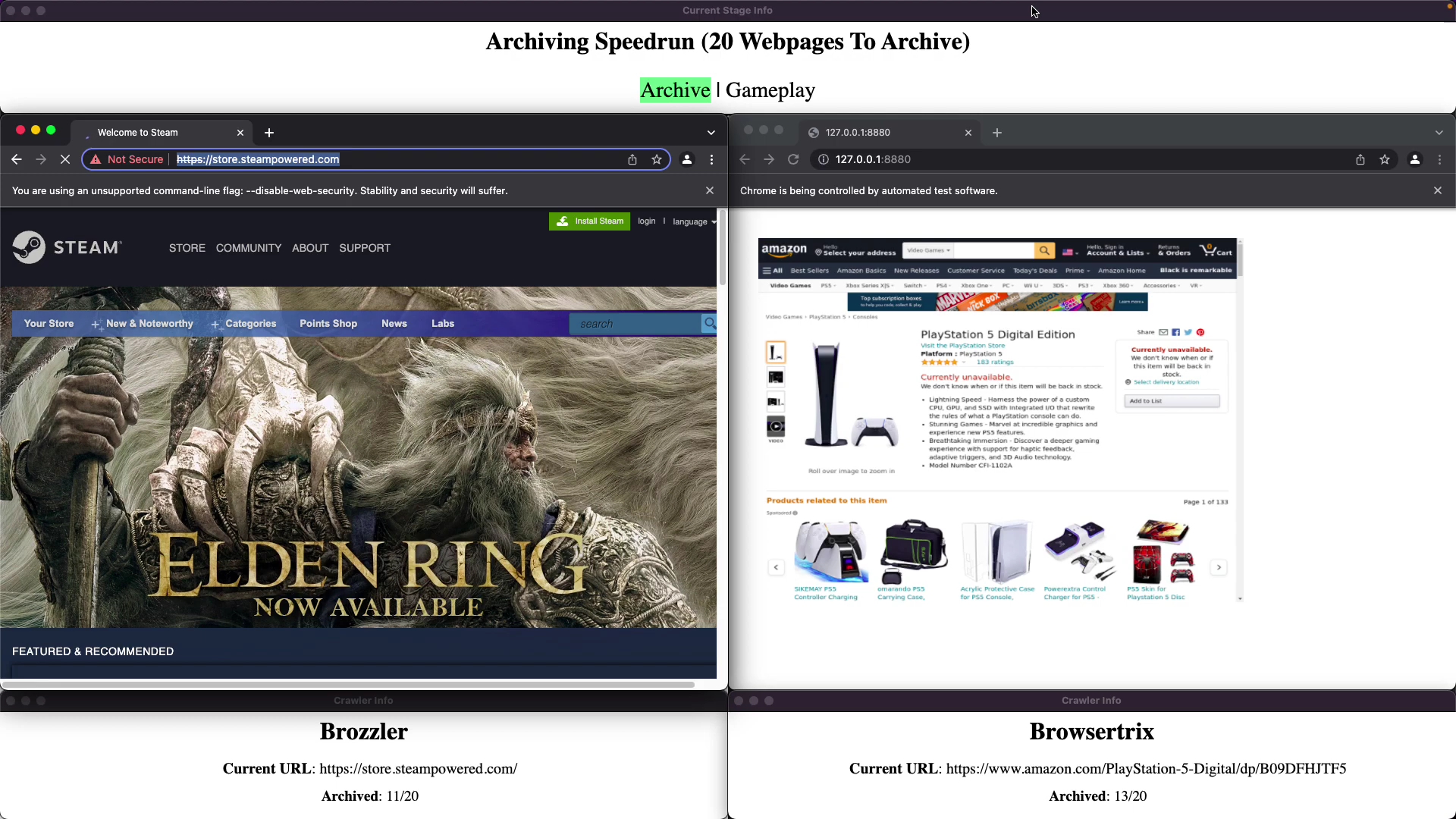}
    \caption{A screenshot that was taken during a web archiving speedrun (video: \href{https://www.youtube.com/watch?v=s2IQERR5V4k}{https://youtu.be/s2IQERR5V4k}) between Brozzler (crawler on the left) and Browsertrix (crawler on the right)}
    \label{fig:speedrun_example}
\end{figure}

We have gamified the web archiving process by applying the gaming concept of speedruns to the web archiving process during our live streams. We created live streams \cite{webArchivingPlaylist-yt22} that showed a competition between two crawlers where the winner of the competition was determined by which crawler finished archiving the given set of seed URIs first. Brozzler and Browsertrix Crawler are the first two crawlers we set up to run in this way and we plan on using other web crawlers in future web archiving live streams. Figure \ref{fig:speedrun_example} is a screenshot from a web archiving speedrun \cite{archivingSpeedrun-yt22} between Brozzler and Browsertrix. During the web archiving live stream, Selenium was used to set up most of the browsers needed for the live stream. The only browser that is not automated by Selenium is Brozzler's browser, because Brozzler can automatically open a browser when it is archiving a web page.

The results of ten web archiving speedruns are shown in Table \ref{table:speedrunTable}. The same set of 20 URIs \cite{uriSets-gh22} was used for all ten speedruns. For Table \ref{table:speedrunTable} the speedrun times for Brozzler and Browsertrix were calculated by taking the finished timestamp for the crawler and then subtracting the beginning of the round timestamp. Brozzler won each of the ten rounds. Browsertrix was able to reach the halfway point (10 URLs archived) faster than Brozzler, but did not win the matches, because Brozzler was faster at archiving the Epic Game Store’s web page and multiple Amazon web pages like the web page shown on the right side of Figure \ref{fig:speedrun_example}. When archiving the Amazon web pages, Browsertrix archived more HTML, JavaScript, CSS, and JSON files than Brozzler, which could be the reason why it took Browsertrix longer to archive these web pages.

\begin{table}[htbp]
\begin{tabular}{|
>{\columncolor[HTML]{FFFFFF}}c |
>{\columncolor[HTML]{FFFFFF}}r |
>{\columncolor[HTML]{FFFFFF}}r |
>{\columncolor[HTML]{FFFFFF}}l |}
\hline
\cellcolor[HTML]{EFEFEF}\textbf{Round} & \multicolumn{1}{c|}{\cellcolor[HTML]{EFEFEF}\textbf{\begin{tabular}[c]{@{}c@{}}Brozzler \\ speedrun time\end{tabular}}} & \multicolumn{1}{c|}{\cellcolor[HTML]{EFEFEF}\textbf{\begin{tabular}[c]{@{}c@{}}Browsertrix \\ speedrun time\end{tabular}}} & \multicolumn{1}{c|}{\cellcolor[HTML]{EFEFEF}\textbf{Video}} \\ \hline
1                                      & \textbf{0:19:16}                                                                                                        & 0:21:52                                                                                                                    & {\url{youtu.be/s2IQERR5V4k}}                          \\ \hline
2                                      & \textbf{0:19:16}                                                                                                        & 0:21:13                                                                                                                    & {\url{youtu.be/wELcdpiOZ74}}                          \\ \hline
3                                      & \textbf{0:19:38}                                                                                                        & 0:22:22                                                                                                                    & {\url{youtu.be/57Z-NxA6Exc}}                          \\ \hline
4                                      & \textbf{0:20:17}                                                                                                        & 0:23:28                                                                                                                    & {\url{youtu.be/WkH1MHfDZMk}}                          \\ \hline
5                                      & \textbf{0:18:47}                                                                                                        & 0:19:15                                                                                                                    & {\url{youtu.be/KooL35aPnmc}}                          \\ \hline
6                                      & \textbf{0:19:49}                                                                                                        & 0:22:16                                                                                                                    & {\url{youtu.be/SHohbwJ07\_Q}}                         \\ \hline
7                                      & \textbf{0:19:16}                                                                                                        & 0:21:59                                                                                                                    & {\url{youtu.be/k9kz59go-88}}                          \\ \hline
8                                      & \textbf{0:19:37}                                                                                                        & 0:22:18                                                                                                                    & {\url{youtu.be/lVX7CiUU1S0}}                          \\ \hline
9                                      & \textbf{0:19:53}                                                                                                        & 0:22:04                                                                                                                    & {\url{youtu.be/zipJR0SwIV0}}                          \\ \hline
10                                     & \textbf{0:19:47}                                                                                                        & 0:21:54                                                                                                                    & {\url{youtu.be/7pcX0biF-Xc}}                          \\ \hline
\textbf{Average speedrun time}         & \textbf{0:19:34}                                                                                                        & 0:21:52                                                                                                                    &                                                             \\ \hline
\end{tabular}
\caption{Results from running 10 archiving speedruns with Brozzler and Browsertrix on the same set of 20 URIs} 
\label{table:speedrunTable} 
\end{table}

\section{Using Web Archiving To Inform Gameplay} \label{Using Web Archiving To Inform Gameplay}
\begin{figure}[htbp]
    \centering
    \includegraphics[scale=0.175]{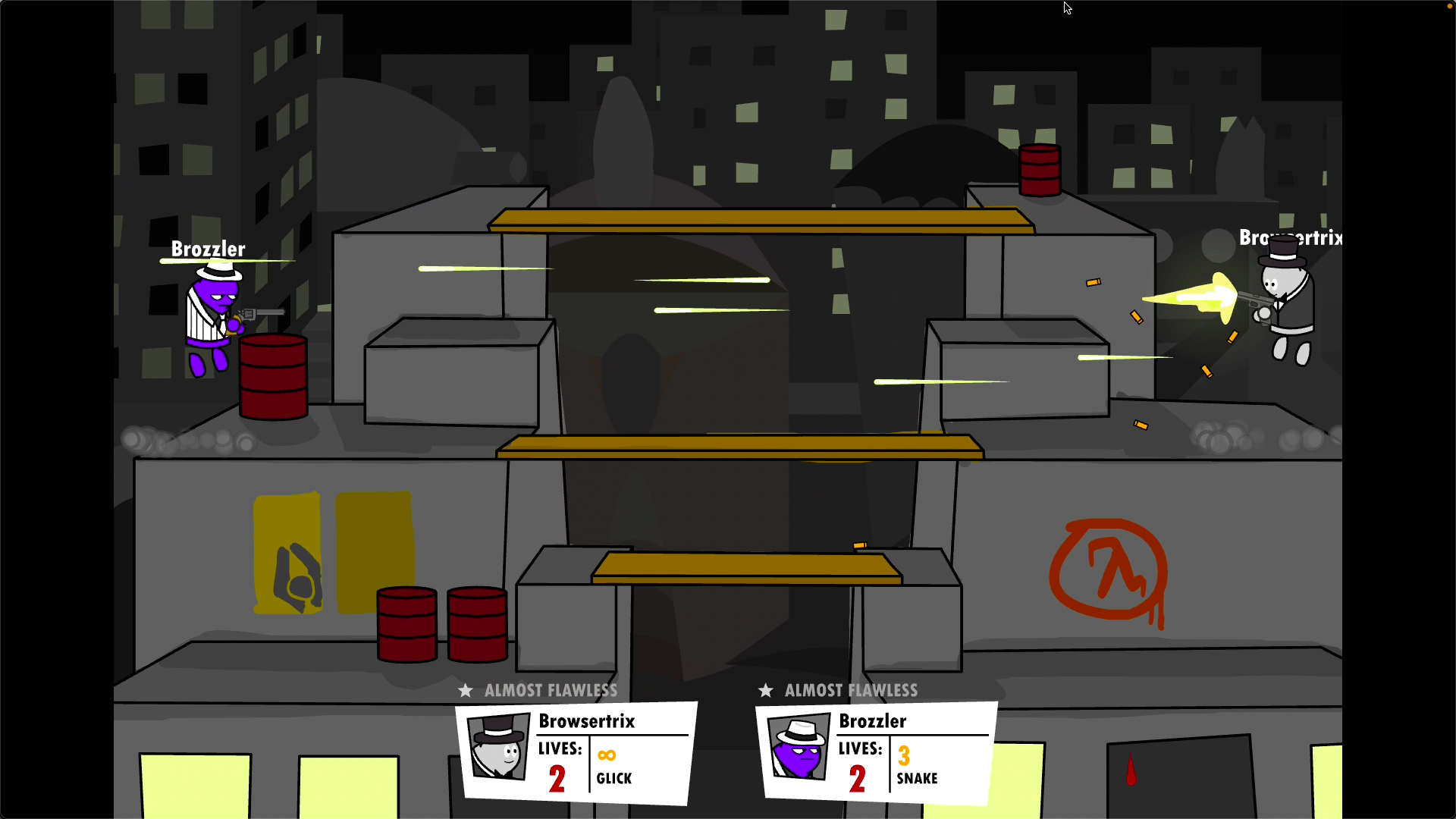}
    \caption{Automated gameplay from Gun Mayhem 2 More Mayhem where the in-game stats for the players were determined based on each crawler's performance.}
    \label{fig:gunMayhem2}
\end{figure}

\begin{figure}[htbp]
    \centering
    \includegraphics[scale=0.175]{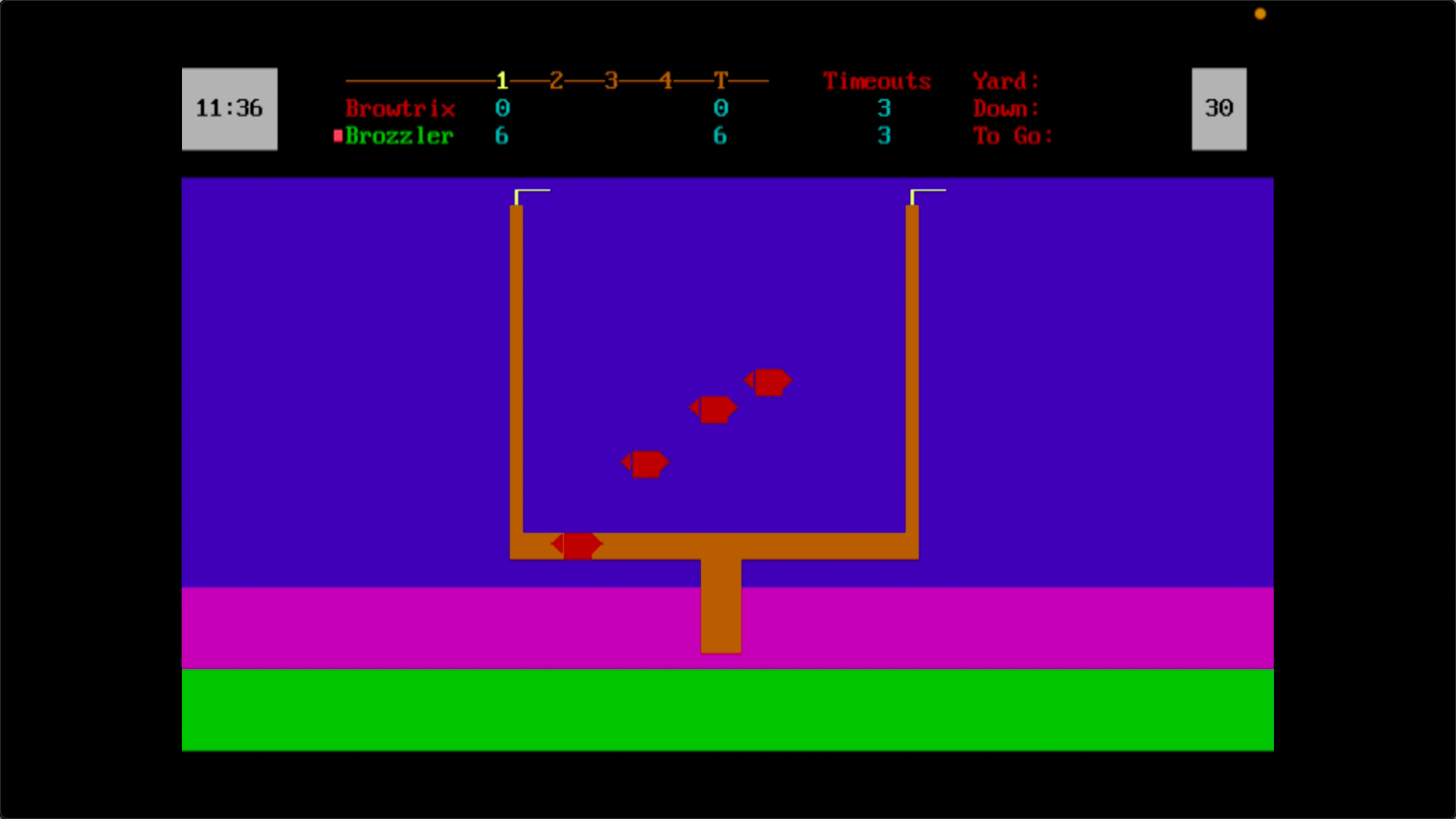}
    \caption{Automated gameplay from NFL Challenge where the in-game stats for the players were determined based on each crawler's performance.}
    \label{fig:nflChallenge}
\end{figure}

When integrating gameplay with web archiving, we used the results of a crawling session to influence the selections that were made for the game. Attributes from the crawling session can be mapped to the settings that were used in the game. Some example attributes that can be used are how long it takes the crawler to finish archiving a set of URLs and the number of resources archived. Determining the quality of the replayed web page should also be considered and approaches like Brunelle’s memento damage \cite{brunelle-jcdl15, brunelle-odu16} and Kiesel's reproduction quality measure \cite{kiesel-acm18} can be used to determine the quality of the archived web page. After we received the results of a crawling session (crawler speed), we used a GUI testing tool to interact with a video game so that we could adjust the settings in the game. For browser gaming, like Flash, HTML5, and cloud gaming, we used Selenium which is a cross-platform tool for browser automation. For locally installed games, like games downloaded from stores like Steam and Epic Game Store, we used Appium to automate desktop applications. Figures \ref{fig:gunMayhem2} and \ref{fig:nflChallenge} are screenshots of matches between Brozzler and Browsertrix where the in-game stats for the  characters or team in the game were determined by each crawler's performance. 
%Figure \ref{fig:gameScreenshots} shows screenshots of matches between Brozzler and Browsertrix where the in-game stats for the  characters or team in the game were determined by each crawler's performance. 

%\begin{figure}[htbp] \centering \begin{subfigure}[b]{0.45\textwidth} \centering \includegraphics[width=\textwidth]{Images/Gun_Mayhem_2_match.png} \caption{Gun Mayhem 2 More Mayhem} \label{fig:gunMayhem2} \end{subfigure} \hfill \begin{subfigure}[b]{0.45\textwidth} \centering \includegraphics[width=\textwidth]{Images/extra_point.png} \caption{NFL Challenge} \label{fig:nflChallenge} \end{subfigure} \caption{Automated gameplay where the in-game stats for the players were determined based on each crawler's performance.} \label{fig:gameScreenshots} \end{figure}
%
%\begin{figure}[htbp] \centering \begin{subfigure}[b]{\textwidth} \centering \includegraphics[width=\textwidth]{Images/Gun_Mayhem_2_match.png} \caption{Gun Mayhem 2 More Mayhem} \label{fig:gunMayhem2} \end{subfigure} \hfill \begin{subfigure}[b]{\textwidth} \centering \includegraphics[width=\textwidth]{Images/extra_point.png} \caption{NFL Challenge} \label{fig:nflChallenge} \end{subfigure} \caption{Automated gameplay where the in-game stats for the players were determined based on each crawler's performance.} \label{fig:gameScreenshots} \end{figure}
%

\begin{figure}[htbp]
     \centering
     \begin{subfigure}[b]{0.45\textwidth}
        \centering
         \includegraphics[scale=0.40]{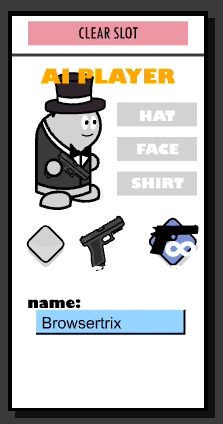}
         \caption{Fastest Web Archive Crawler}
         \label{fig:gm2Fast}
     \end{subfigure}
     \hfill
     \begin{subfigure}[b]{0.45\textwidth}
        \centering
         \includegraphics[scale=0.40]{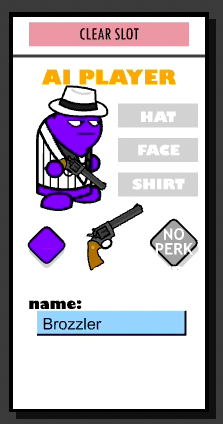}
         \caption{Slowest Web Archive Crawler}
         \label{fig:gm2Slow}
     \end{subfigure}
     \caption{The character selections made for the fastest and the slowest web crawlers.}
     \label{fig:gm2traits}
     %\hfill
\end{figure}

The first game we used for a gaming live stream was Gun Mayhem 2 More Mayhem (Figure \ref{fig:gunMayhem2}).  Gun Mayhem 2 More Mayhem is a platform fighting game where the goal is to knock the opponent off the stage. When a player gets knocked off the stage, they lose a life. The winner of the match will be the last player left on the stage. Gun Mayhem 2 More Mayhem is a Flash game that is played in a web browser. Selenium was used to automate this game since it is a browser game. When selecting the settings in the game, it was not possible to simulate key presses to change the settings so we had to simulate mouse clicks on the game’s HTML canvas to make the in-game selections. In the Gun Mayhem 2 More Mayhem demo \cite{gm2Demo-yt22}, the crawler's speed was used to determine which perk to use and the gun to use. Some example perks are infinite ammo, triple jump, and no recoil when firing a gun. The fastest crawler used the fastest gun and was given an infinite ammo perk (Figure \ref{fig:gm2Fast}). The slowest crawler used the slowest gun and did not get a perk (Figure \ref{fig:gm2Slow}).

\begin{figure}[htbp]
     \centering
     \begin{subfigure}[b]{0.45\textwidth}
         \centering
         \includegraphics[width=\textwidth]{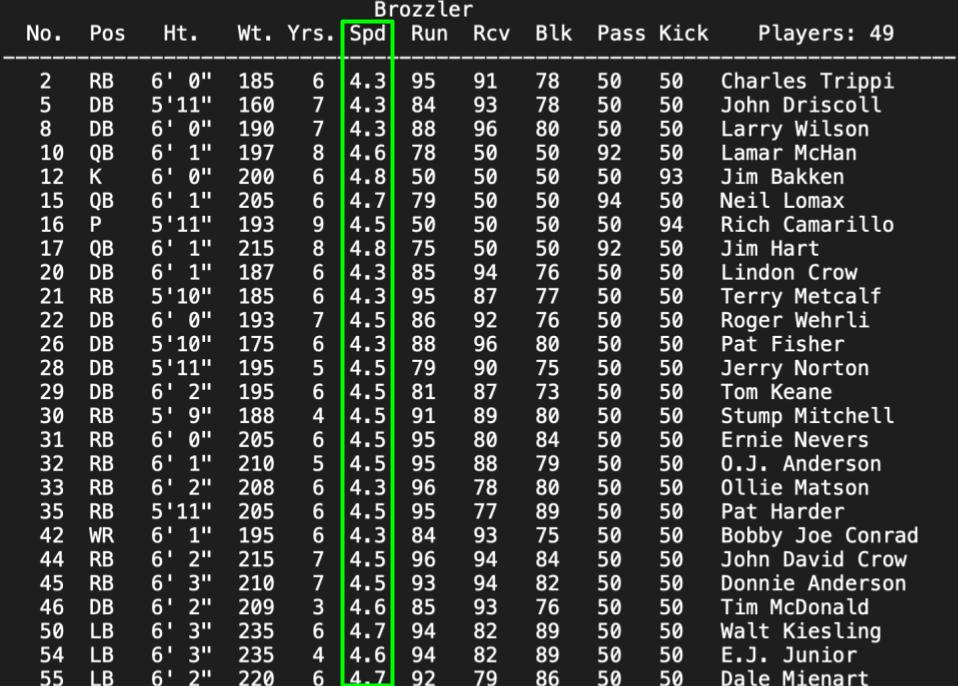}
         \caption{Fastest Web Archive Crawler}
         \label{fig:nflcFast}
     \end{subfigure}
     \hfill
     \begin{subfigure}[b]{0.45\textwidth}
         \centering
         \includegraphics[width=\textwidth]{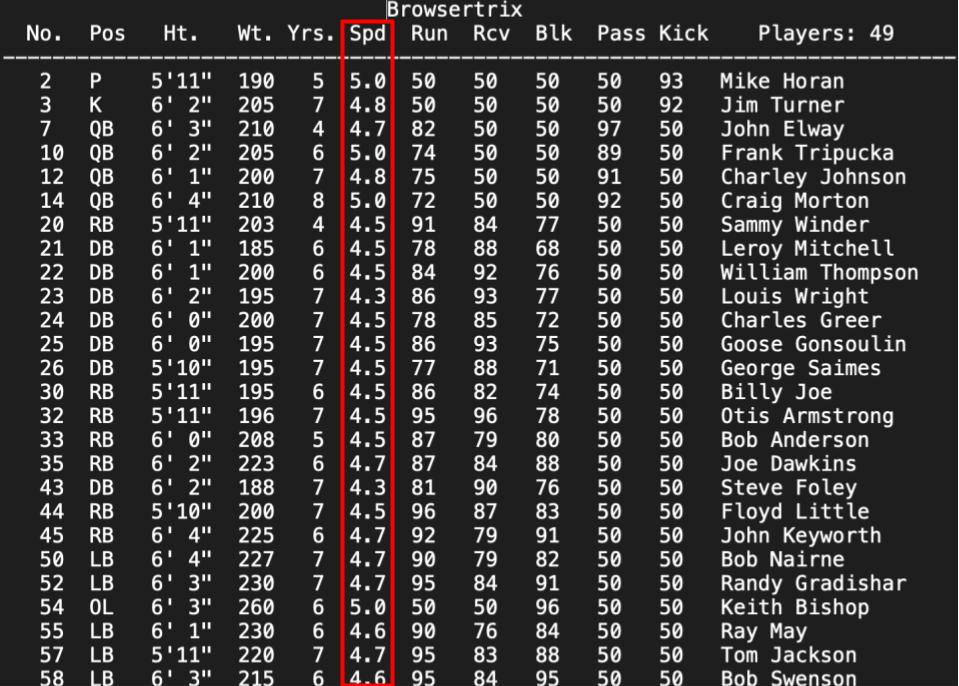}
         \caption{Slowest Web Archive Crawler}
         \label{fig:nflcSlow}
     \end{subfigure}
     \caption{The player attributes for the teams associated with the fastest and slowest web crawlers. The speed ratings are the times for the 40-yard dash, so the lower numbers are faster.}
     \label{fig:nflctraits}
\end{figure}

The second game that was used during our gaming live streams is NFL Challenge. NFL Challenge (Figure \ref{fig:nflChallenge}) is an NFL football simulator that was released in 1985 and was popular during the 1980s. The performance of a team is based on the player attributes that are stored in editable text files. It is possible to change the stats for the players, like the speed, passing, and kicking ratings, and it is possible to change the name of the team and the players on the team. This customization allows us to rename the team’s name to the name of the web crawler and to rename the players of the team to the names of the contributors of the tool. NFL Challenge is an MS-DOS game that can be played with an emulator named DOSBox. Appium was used to automate the game since NFL Challenge is a locally installed game. Since NFL Challenge is a game that is executed in a command-line interface only key presses were simulated by Appium when making selections in the game. In the NFL Challenge demo \cite{nflcDemo-yt22}, the fastest crawler would get the team with the fastest players based on the players’ speed attribute (Figure \ref{fig:nflcFast}) and the other crawler would get the team with the slowest players (Figure \ref{fig:nflcSlow}).

\section{Future Work}
We plan on making several improvements to the live streams. The web archiving live stream needs to include a mode that shows a summary of the web archiving and replay performance results, and more gaming concepts will be applied to the web archiving live streams. Also, for the web archiving live stream, we need to evaluate if the live streams are entertaining and determine the benefits of using live streams to add transparency to the web archiving process. The gaming live streams and web archiving live streams will be updated so that both live streams will run at the same time. We will also automate more example games that are from different gaming platforms and game genres.

%%% | Adding Results Mode To The Web Archiving Live Streams
Our web archiving live stream currently can show two crawlers archiving web pages (archive mode) and the playback of the archived web pages (replay mode). In future live streams we will add a results mode so that viewers can see a summary of the web archiving and replay performance results and see which crawler won the round based on the number of points the crawler received.
%%% | Adding Replay To The Web Archiving Live Streams

%%% | Running Web Archiving Live Stream and Gaming Focused Live Stream At The Same Time
When running our web archiving live stream and gaming live stream, they currently cannot be run at the same time. The demos we have so far begin with the crawling session and then switch to the gaming session. We plan on having two separate live streams running at the same time where one live stream shows the crawling and replay sessions and the other live stream shows the gameplay that is influenced by the crawlers' performance. 

%%% | Map More Crawler Performance Attributes To Video Game Settings
The web archiving live stream will use more than the speed of a web archive crawler when determining the crawler’s performance, such as using metrics from Brunelle's \cite{brunelle-jcdl15, brunelle-odu16} memento damage algorithm which is used to measure the replay quality of archived web pages. During future web archiving live streams, we will also evaluate and compare the capture and playback of web pages archived by different web archives and archiving tools like the Wayback Machine, archive.today, and Arquivo.pt.

%%% | Apply More Gaming Concepts To The Web Archiving Process
We plan on applying more gaming concepts to the web archiving live streams. So far, we have only applied the gaming concept of speedruns to web archiving live streams. The next gaming concept we will apply will be similar to an arcade mode in a fighting video game or a tournament. In arcade mode, one character is selected to fight against all of the other characters. Each match in the arcade mode would only have two characters fighting against each other and the arcade mode ends when all of the other characters have been defeated. For the tournament, there would be individual matches between crawlers and the winner of the final match will win the tournament.

%%% | Evaluate If Gamification of The Web Archiving Process Makes Web Archiving Live Streams Entertaining
We have not performed an evaluation yet, because conventional metrics for user engagement with a video like the number of views, likes and dislikes, and average view duration may not be ideal for determining if applying gamification to the web archiving process will make our videos entertaining and determining the benefits of using live streams to add transparency to the web archiving process. In the future, we will evaluate if applying gaming concepts to the web archiving process is enough to make the web archiving live streams entertaining to watch. We will also evaluate if there are any benefits of adding transparency to the web archiving process by streaming the web archiving sessions and replay sessions.

%%% | Integrate More Games From Different Gaming Platforms, Operating Systems, and Gaming Genres
We will update the gaming live streams so that they can support more games and games from different genres. The games that we supported so far are multiplayer games. We will also try to automate single-player games where the in-game character for each crawler can compete to see which player gets the highest score on a level or which player finishes the level the fastest. For games that allow creating a level or game world, we would like to use what happens during a crawling session to determine how the level is created. If the crawler was not able to archive most of the resources, then more enemies or obstacles could be placed on the level to make it more difficult to complete the level. Some games that we will try to automate include: Rocket League, Brawhalla, Quake, and DOTA 2. When the scripts for the gaming live stream are ready to be released publicly it will also be possible for anyone to add support for more games that can be automated.  

We will also have longer runs for the gaming live streams so that a campaign or season in a game can be completed. A campaign is a game mode where the same characters can play a continuing story until it is completed. A season for the gaming live streams will be like a season for a sport where there are a certain number of matches that must be completed for each team during a simulated year and a playoff tournament that ends with a championship match.

\section{Conclusions}
We are developing a proof of concept that involves gamification of the web archiving process and integrating video games with web archiving. We have integrated gaming and web archiving so that web archiving can be more entertaining to watch and enjoyed like a spectator sport. We have applied the gaming concept of a speedrun to the web archiving process by having a competition between two crawlers where the crawler that finished archiving the set of seed URIs first would be the winner. We have also created automated web archiving and gaming live streams where the web archiving performance of web crawlers from the web archiving live streams were used to determine the capabilities of the characters inside of the Gun Mayhem 2 More Mayhem and NFL Challenge video games that were played during the gaming live streams. Our contribution is making the web archiving process more transparent to third parties by live streaming the crawling of web pages during the web archiving session and the playback of the archived web pages during the replay session.

\subsection*{Acknowledgments}
The Game Walkthroughs and Web Archiving project has received a seed grant from the International Internet Preservation Consortium (IIPC). We thank the IIPC for supporting this work.

\bibliographystyle{splncs04}
\bibliography{refs}
%\end{thebibliography}
\end{document}